\documentstyle[emulateapj]{article}

\slugcomment{Submitted to the Astrophysical Journal}

\lefthead{Hamilton \& Fesen}
\righthead{\ion{Fe}{2} Image of SN~1885}

\newcommand{\be}{\begin{equation}}
\newcommand{\ee}{\end{equation}}
\newcommand{\ba}{\begin{eqnarray}}
\newcommand{\ea}{\end{eqnarray}}

\newcommand{\Ca}{{\rm Ca}}
\newcommand{\CaI}{{\rm Ca I}}
\newcommand{\CaII}{{\rm Ca II}}
\newcommand{\Fe}{{\rm Fe}}
\newcommand{\FeI}{{\rm Fe I}}
\newcommand{\FeII}{{\rm Fe II}}
\newcommand{\Msun}{\mbox{${\rm M}_{\sun}$}}
\newcommand{\Z}{{\rm Z}}
\newcommand{\ZI}{{\rm Z I}}
\newcommand{\ZII}{{\rm Z II}}

\newcommand{\cm}{{\rm cm}}
\newcommand{\cts}{{\rm cts}}
\newcommand{\erg}{{\rm erg}}

\newcommand{\kms}{{\rm km}\,{\rm s}^{-1}}

\newcommand{\pc}{{\rm pc}}

\newcommand{\pix}{{\rm pix}}
\newcommand{\s}{{\rm s}}
\newcommand{\yr}{{\rm yr}}

\newcommand{\NoteToEd}[1]{}

\newcommand{\iontable}{
    \begin{deluxetable}{ccc}
    \tablewidth{0pt}
    \tablecaption{Photoionization timescales
    \label{iontable}}
    \tablehead{
\colhead{Element} & \colhead{Photoionization time\tablenotemark{a}} & \colhead{\ion{Z}{2}/\ion{Z}{1}\tablenotemark{b}} \\
& (yr) &
}
    \startdata
C & $3.9^{+4.7}_{-0.4}$ & $300^{+4000}_{-140}$ \\
O\tablenotemark{c} & long & $0$ \\
Mg & $44^{+53}_{-4}$ & $0.7^{+0.4}_{-0.1}$ \\
Al & $1.0^{+1.1}_{-0.1}$ & $\sim 10^{10}$ \\
Si & $0.7^{+0.9}_{-0.1}$ & $\sim 10^{13}$ \\
S & $1.3^{+1.5}_{-0.1}$ & $\sim 10^8$ \\
Ar\tablenotemark{c} & long & $0$ \\
Ca & $8^{+10}_{-1}$ & $16^{+42}_{-5}$ \\
Ti & $9^{+12}_{-1}$ & $10^{+20}_{-3}$ \\
V & $23^{+29}_{-2}$ & $1.6^{+1.4}_{-0.3}$ \\
Cr & $4^{+4}_{-1}$ & $600^{+10000}_{-300}$ \\
Mn & $111^{+135}_{-9}$ & $0.23^{+0.11}_{-0.03}$ \\
Fe & $10^{+11}_{-1}$ & $10^{+19}_{-3}$ \\
Co & $63^{+76}_{-6}$ & $0.44^{+0.24}_{-0.07}$ \\
Ni & $35^{+42}_{-3}$ & $0.9^{+0.7}_{-0.1}$ \\
Cu & $9^{+10}_{-1}$ & $13^{+33}_{-4}$ \\
Zn & $62^{+75}_{-6}$ & $0.44^{+0.25}_{-0.06}$ \\
    \enddata
    \tablenotetext{a}{
Ionization timescale in optically thin limit;
actual ionization timescale is probably longer.
}
    \tablenotetext{b}{
Predicted ratio of singly-ionized to neutral abundance.
}
    \tablenotetext{c}{
Ionization potentials of these elements exceed the Lyman limit.
}
    \end{deluxetable}
}

\newcommand{\imagefig}{
    \begin{figure*}
    \begin{center}
    \leavevmode
    \epsfbox{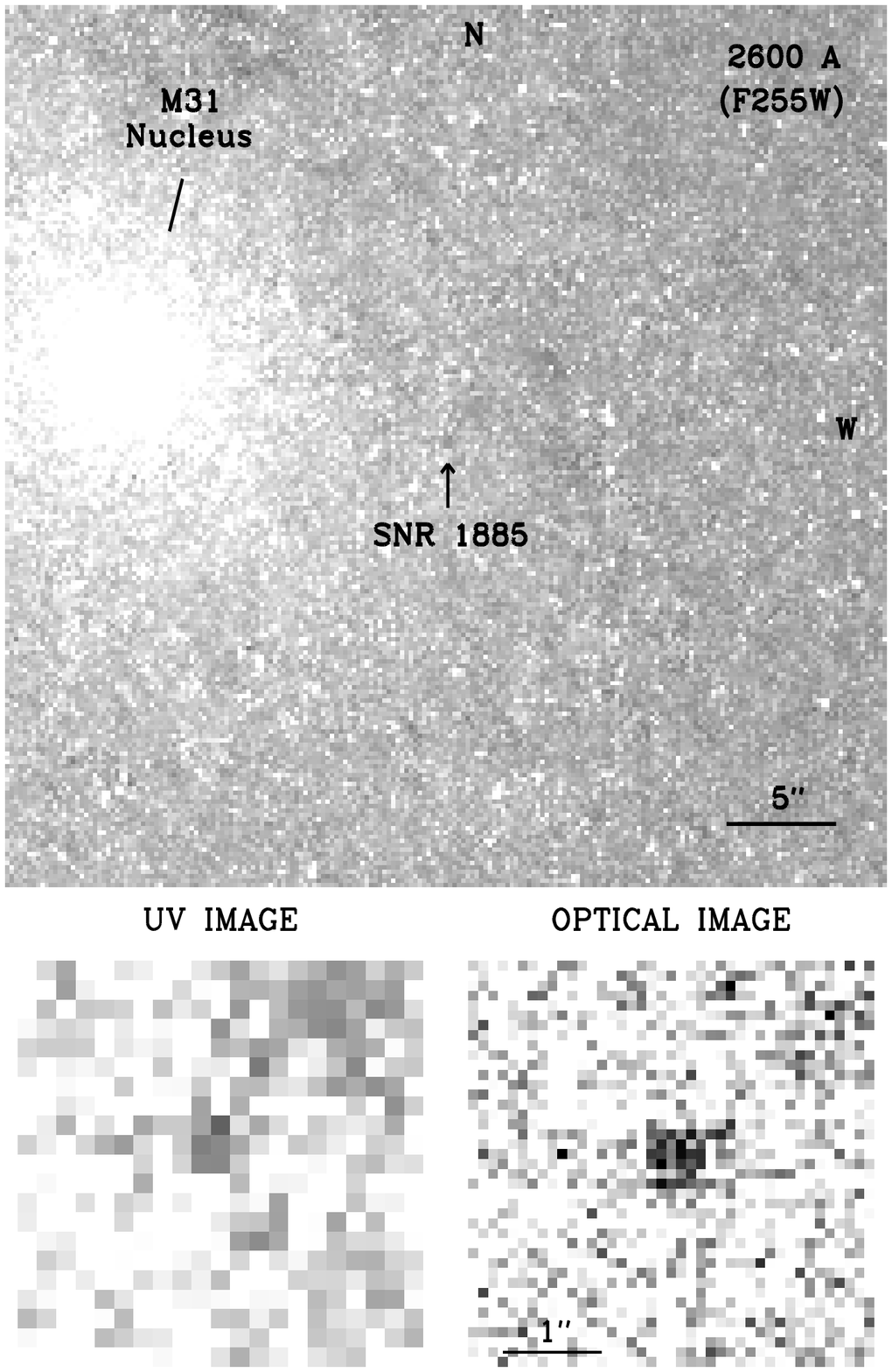}
    \end{center}
    \caption[1]{\small
    \imagecaption
    }
    \end{figure*}
}
\newcommand{\imagecaption}{
WF2 F255W image of SNR~1885 in the bulge of M31.
Some obscuration by dust is visible,
notably along a broad lane from top left to bottom right,
somewhat above center.
Bottom panels show close-ups of the region around the position of SNR~1885,
at left in the `\ion{Fe}{2}' F255W filter,
and at right in the `\ion{Ca}{2}' FQUVN-D filter,
at the same scale.
The \ion{Ca}{2} image is from \protect\markcite{FGMH}Paper~1.
SNR~1885 is the black absorbing spot at the center of the close-ups.
The dark patch to the top right in the close-ups is
part of the broad lane of dust visible in the full image.
\label{image}
}

\newcommand{\specuvfig}{
    \begin{figure*}[tb]
	\specuv
    \end{figure*}
}
\newcommand{\specuv}{
    \epsfxsize=3.4in
    \epsfbox{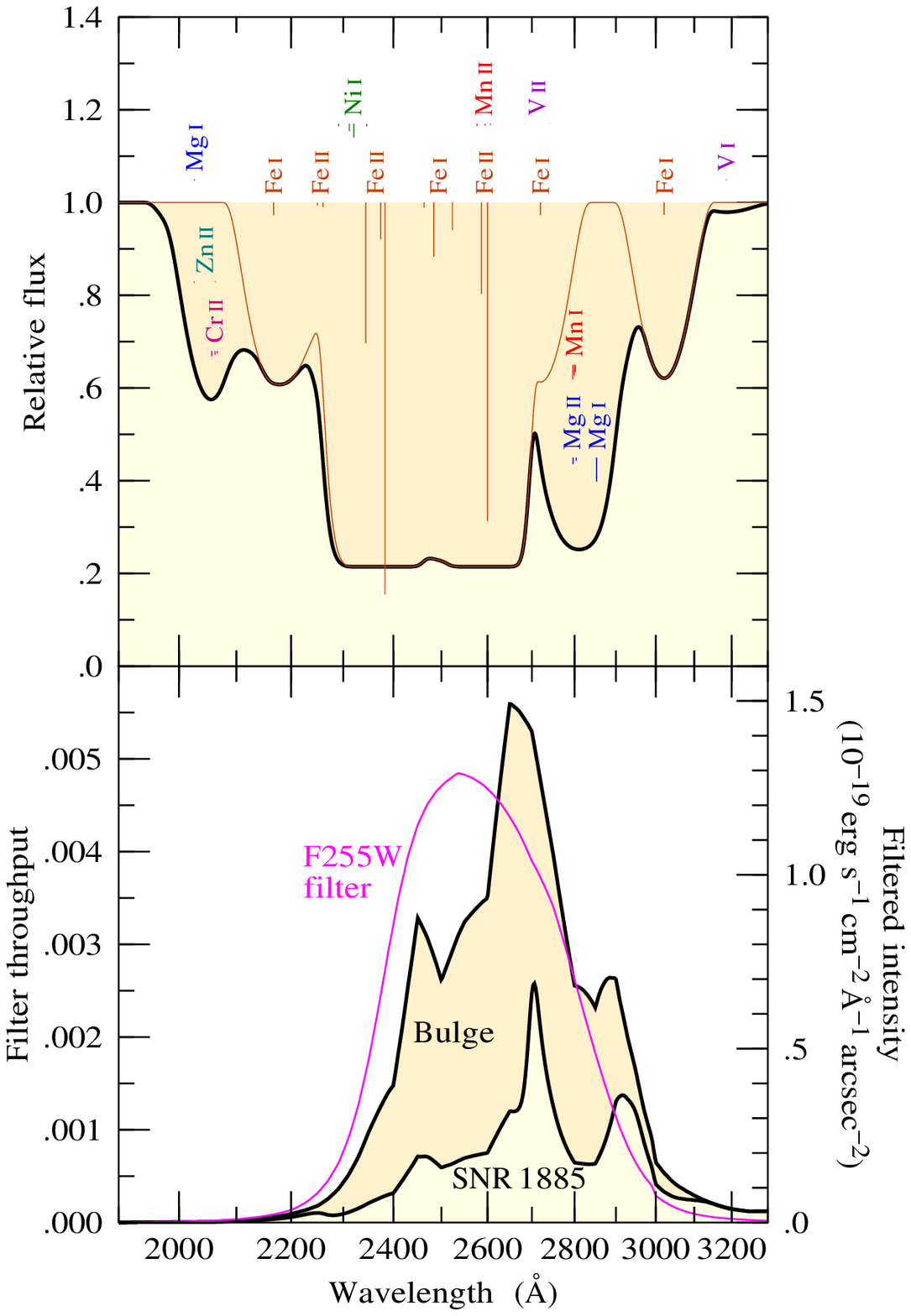}
    \caption[1]{\small
    \specuvfigcaption
    }
}
\newcommand{\specuvfigcaption}{
(Top)
Model UV absorption spectrum over $1900$--$3300 \, {\rm \AA}$
predicted from a model fit to the near-UV 3240--$4780 \, {\rm \AA}$ spectrum
observed with G400H on the FOS (\protect\markcite{FGMH}Paper~1).
Thin line shows the absorption from Fe alone;
thick line includes absorption from all elements.
Vertical lines mark principal resonance lines included in the model.
The lengths of the vertical lines are proportional
to oscillator strength times wavelength times ion abundance for each
resonance line, a measure of the optical depth of the line.
The minimum flux level of $0.21$ is produced by unabsorbed starlight
to the foreground of SNR~1885 in the bulge of M31.
(Bottom)
Thin line is the throughput of the F255W filter with the WF2 chip on WFPC2.
Upper thick line is the expected spectrum from the bulge of M31
as measured by {\it IUE\/} (\protect\cite{Burstein88}),
normalized to the surface brightness at the position of SNR~1885,
and multiplied by the filter throughput.
Lower thick line is the expected absorbed spectrum at SNR~1885,
the product of the filtered bulge spectrum with the absorption spectrum
in the top panel.
The ratio of the area under the SNR~1885 spectrum
to the area under the bulge spectrum, here 0.33,
is the predicted fractional depth of absorption
at SNR~1885's position in the F255W image.
\label{specuv}
}

\begin{document}

\title{An Ultraviolet \ion{Fe}{2} Image of SN~1885 in M31}

\author{Andrew J. S. Hamilton}
\affil{JILA \& Dept. of Astrophysical \& Planetary Sciences,
	U. Colorado, Boulder, CO 80309;
	Andrew.Hamilton@colorado.edu}
\and
\author{Robert A. Fesen}
\affil{6127 Wilder Laboratory, Physics \& Astronomy Department
	Dartmouth College, Hanover, NH 03755;
	fesen@snr.dartmouth.edu}

\begin{abstract}
Ultraviolet imaging of the remnant of Supernova 1885 in M31 with the
{\it Hubble Space Telescope\/} using the F255W filter on the WFPC2
reveals a dark spot of \ion{Fe}{2} absorption at the remnant's known position
in the bulge of M31.
The diameter of the absorbing spot is $0\farcs55 \pm 0\farcs15$,
slightly smaller than, but consistent with,
the $0\farcs70 \pm 0\farcs05$ diameter
measured in the higher quality WFPC2 \ion{Ca}{2} absorption image
previously reported by us.
The measured ratio of flux inside to outside SNR~1885 in the \ion{Fe}{2} image
is $0.24 \pm 0.17$,
consistent with the ratio $0.33 \pm 0.04$
expected on the basis of a model fit to the previously obtained
near-UV FOS spectrum.
The observed depth of \ion{Fe}{2} absorption suggests that \ion{Fe}{2} is
fully saturated, with an iron mass in the range $M_\Fe = 0.1$--$1.0 \, \Msun$.
Besides Fe, ion species \ion{Mg}{1}, \ion{Mg}{2}, and \ion{Mn}{1} 
probably make some contribution to the absorption from the SN~1885 remnant
in the F255W image.
\end{abstract}


\section{Introduction}

SN~1885 (S~Andromedae) in the Andromeda galaxy M31
was the first supernova recorded in another galaxy (\cite{Zwicky}).
The supernova, which occurred just $15 \farcs 6 \pm 0 \farcs 1$
from the central nucleus of M31,
reached a peak $V$ magnitude of 5.85 in August 1885
(\cite{DVC85}).
At the $725 \pm70 \, \pc$ distance of M31,
and allowing for 0.23~mag of extinction,
this corresponds to an absolute magnitude of $M_V = -18.7$ (\cite{vdb94}),
some 0.8 mag fainter than the peak magnitude
$M_V = -19.48 \pm 0.07$ of normal SN~Ia (\cite{Tammann99}).
This, combined with SN~1885's unusually fast light curve
and reddish color near maximum light
(\cite{DVC85}),
point to a subluminous Type~Ia event
similar to SN~1991bg (\cite{Fil97}).

A little over a century later,
the remnant of SN~1885 (SNR~1885) was detected
through a near-UV filter ($3900 \pm 100 \, {\rm \AA}$)
as a spot of absorption
silhouetted against the starlight of M31's central bulge
(\cite{FHS}).
Fesen et al.\ attributed the absorption to the resonance line of
\ion{Fe}{1} $3860 \, {\rm \AA}$,
consistent with the expected presence of a large mass of
iron in a Type~Ia supernova.

Subsequent near-UV WFPC2 imaging and FOS spectroscopy with
the {\it Hubble Space Telescope\/} ({\it HST})
revealed that the principal source of absorption is
not \ion{Fe}{1}, but rather \ion{Ca}{2} H \& K,
freely expanding at velocities up to $13\,100 \pm 1500 \, \kms$
(\cite{FGMH}, hereafter Paper~1).
In addition to strong \ion{Ca}{2} H \& K absorption,
the FOS spectrum showed similarly broad but weaker absorption
from \ion{Ca}{1} $4227 \, {\rm \AA}$
and \ion{Fe}{1} $3720 \, {\rm \AA}$ ($5v$),
and possibly $3441 \, {\rm \AA}$ ($6v$)
and $3860 \, {\rm \AA}$ ($4v$).

The \ion{Fe}{1} absorption observed in the FOS spectrum
implies a mass of $M_\FeI = 0.013^{+0.010}_{-0.005} \, \Msun$
in the ejecta of SNR~1885.
The observed relative strengths of the \ion{Ca}{2} and \ion{Ca}{1} lines
indicate that calcium is mostly singly ionized, with
$M_\CaII/M_\CaI = 16^{+42}_{-5}$,
the large upward uncertainty
reflecting the near saturation of the \ion{Ca}{2} H \& K feature.
If the ionization state of iron is similar to that of calcium,
with $M_\FeII/M_\FeI \approx 10$--$50$,
then the corresponding \ion{Fe}{2} mass is
$M_\FeII \approx 0.1$--$0.7 \, \Msun$.
Such a large mass of iron is consistent with what is expected
in normal or subluminous Type~Ia supernovae
(\cite{HK96}; \cite{HWT98}; \cite{NTY84}; \cite{N97}; \cite{WW94}; \cite{W97}).
Thus on both observational and theoretical grounds
there is good reason to expect that the remnant of SN~1885
should show strong \ion{Fe}{2} resonance line absorption.

In this paper we report the detection of \ion{Fe}{2} absorption
in a UV image of SNR~1885 obtained with the WFPC2 on {\it HST}.

\section{Observations}

\NoteToEd{
EDITOR: PLACE FIGURE 1 ON PAGE 2 OF THE PUBLISHED PAPER, TAKING UP THE ENTIRE 
PAGE.
}

\imagefig

\subsection{Images}

The strongest resonance lines of \ion{Fe}{2}
are the 2600, $2587 \, {\rm \AA}$ ($1uv$)
and 2382, 2344, $2374 \, {\rm \AA}$ ($2,3uv$) complexes.
A model fit to the near-UV spectrum reported in \markcite{FGMH}Paper~1
predicts that these \ion{Fe}{2} resonance lines should
form a broad, deep, blended profile that is fortuitously
well matched to the WFPC2 F255W ($2597 \pm 200 \, {\rm \AA}$) filter
on {\it HST}.

Three UV exposures were taken with the WFPC2 and F255W filter
over 3 orbits on 16 Feb 1999.
The bulge of M31, though bright enough to see with the naked eye in the visible,
is quite faint in the UV,
and special measures were taken to ensure detection of SNR~1885.
While sky brightness was negligible (about 0.01 of the signal),
both readout noise and dark counts were significant.
Dark counts were reduced by centering SNR~1885 in the WF2 chip,
which has the lowest dark count of the WF and PC chips.
Readout noise was reduced by minimizing the number of readouts,
which was accomplished by $2 \times 2$ on-chip binning (AREA mode),
and by extending each exposure over a full orbit, 2700~s each.
Finally,
the effect of hot and cold pixels was mitigated
by dithering the three images along a diagonal line,
by two binned pixels (4 unbinned pixels)
in each of the horizontal and vertical directions.

The long exposures increased the risk of contamination by cosmic rays,
but this risk was considered acceptable in the interest of reducing noise.
Cosmic rays were removed by applying the {\tt crrej\/} routine
in {\it STSDAS\/} to the three exposures.
Approximately 10\% of the binned pixels in each 2700~s exposure were affected
by cosmic rays.
Cold pixels
were removed by applying the {\tt cosmicray\/} routine
in {\it IRAF\/} to the negative of the cosmic-ray-removed WF2 image.

Figure~\ref{image} shows the resulting cleaned, coadded WF2 image.
SNR~1885 shows up as a dark spot of \ion{Fe}{2} absorption.
Since the absorbing region at SNR~1885's position was partially contaminated
by cosmic rays in both the first and third images,
the pattern of absorption visible in the UV close-up in Figure~\ref{image}
is determined to a considerable degree by the second image.
We estimate the diameter of the dark \ion{Fe}{2} spot to be
$0\farcs55 \pm 0\farcs15$,
slightly smaller than, but consistent with,
the diameter $0\farcs70 \pm 0\farcs05$ of the \ion{Ca}{2} spot
measured in \markcite{FGMH}Paper~1.
The position of the dark spot is consistent with
(within one binned pixel of)
that measured from the higher quality \ion{Ca}{2} WFPC2 image
of \markcite{FGMH}Paper~1,
which was
$15\farcs04 \pm 0\farcs1$ west and $4\farcs1 \pm 0\farcs1$ south
of the nucleus of M31.

The cleaned image shown in Figure~\ref{image}
shows of the order of a hundred point sources.
These point sources appeared only if the 3 exposures were correctly registered.
If instead one or more exposures were misaligned,
then most of the point sources disappeared,
being rejected as cosmic rays by {\tt crrej}.
Visual inspection in several cases confirmed that the point sources
that survive screening by {\tt crrej\/} occur in all three exposures,
and look like stars on each exposure.
We therefore conclude that,
while a handful of the point sources may be cosmic ray artifacts,
the majority ($\ga 90\%$) of them are real stars.

Most of the
stars in the \ion{Fe}{2} ($\sim 2600 \, {\rm \AA}$) image
are not apparent in the
\ion{Ca}{2} ($\sim 3900 \, {\rm \AA}$) image from \markcite{FGMH}Paper~1,
although the mottled appearance of the \ion{Ca}{2} image suggests
incipient resolution into stars (\cite{Lauer98}).
That resolved stars appear only at shorter wavelengths is consistent with
previous UV imaging of the bulge of M31 by
Bertola et al.\ \markcite{Bertola95}(1995),
Brown et al.\ \markcite{Brown98}(1998),
and Lauer et al.\ \markcite{Lauer98}(1998).
Current observational evidence and theoretical ideas,
reviewed by O'Connell \markcite{OConnell99}(1999),
suggest that the UV ($\la 2500 \, {\rm \AA}$) light observed in old populations
such as the bulge of Andromeda is dominated by low-mass, thin-envelope stars
in extreme (hot) horizontal branch and subsequent phases of evolution.
The more luminous UV-bright stars, such as those observed here,
are undergoing hydrogen- and helium-shell burning in later phases of evolution
following core helium burning on the horizontal branch.
Further discussion of this issue goes beyond the scope of this paper.

\subsection{UV Count Levels for SNR 1885}

In the UV absorption image of SNR~1885,
the observable quantity that can be compared to theoretical expectation
is the fractional depth of absorption produced by the remnant
against background starlight from the bulge of M31.
The fractional depth of absorption follows from three quantities:
(a) the zero-level of counts from dark current plus readout,
(b) the background level of counts from starlight in regions adjacent
to SNR~1885,
and (c) the level of counts in SNR~1885 itself.

The zero-level of counts, from dark current plus readout,
was estimated from averages of counts in the darkest regions of the
cosmic-ray-removed WF2 image.
Measurements
at the centers of the darkest dust lanes
and in regions farthest from the bulge
gave a consistent zero-level of
$27 \pm 1 \, {\rm DN}$ (data numbers) per $2 \times 2$ binned pixel
for the coadded $3 \times 2700 \, \s = 8100 \, \s$ image.
At a gain of $7.12 \, \cts \, {\rm DN}^{-1}$,
this corresponds to a zero-level of $192 \pm 7$ counts per binned pixel.
The uncertainty here is an estimate of the uncertainty in the mean zero-level,
not a measure of the variation in pixel to pixel counts, which is larger.
The measured zero-level agrees well with the expected zero-level,
which comprises dark counts of
$(0.0030 \pm 0.0005) \, \cts \  \s^{-1} \pix^{-1} \times 4 \, \pix
\times 3 ( 2700 + 120) \, \s
= 102 \pm 17 \, \cts$
(the uncertainty is the systematic variation
in the dark current of the WF2 chip,
while the $120 \, \s$ added to each $2700 \, \s$ exposure is the unexposed
dark time),
plus readout counts of $3 \times 5.51^2 = 91 \, \cts$,
for a total expected zero-level of $193 \pm 17 \, \cts$ per binned pixel.

The level of background starlight against which SNR~1885 is seen in absorption
was estimated from clean regions adjacent to SNR~1885.
The average and dispersion of the counts in these regions
was $38 \pm 4 \, {\rm DN}$ per binned pixel,
equivalent to $271 \pm 28 \, \cts$ per binned pixel.
Subtracting the zero-level of $192 \pm 7 \, \cts$
gives a background starlight level of
$79 \pm 29 \, \cts$ per binned pixel.
This translates into a surface brightness of
$2.0 \pm 0.7 \times 10^{-17} \, \erg \, \s^{-1} \cm^{-2} {\rm arcsec}^{-2} {\rm \AA}^{-1}$
at $\sim 2600 \, {\rm \AA}$
in the vicinity of SNR~1885.
This surface brightness is consistent with
the mean surface brightness measured by {\it IUE\/} of
$4.4 \times 10^{-17} \, \erg \, \s^{-1} \cm^{-2} {\rm arcsec}^{-2} {\rm \AA}^{-1}$
at $2600 \, {\rm \AA}$ 
through a $154 \, {\rm arcsec}^2$ racetrack-shaped aperture
centered on the nucleus of M31
(\cite{Burstein88}; see also \cite{Bertola95}; \cite{Brown98}).

Counts in SNR~1885 itself were estimated from the $2 \times 2$
($0\farcs4 \times 0\farcs4$) block of binned pixels
centered at the position measured from the \ion{Ca}{2} image of
\markcite{FGMH}Paper~1.
In the \ion{Fe}{2} image,
the dark absorbing region associated with SNR~1885
appears to extend over a block 2 binned pixels ($0\farcs4$) wide (east-west)
by 3 binned pixels ($0\farcs6$) high (north-south).
However, since the northern 2 binned pixels in the $2 \times 3$ block
lie partially outside the \ion{Ca}{2} absorbing region,
we conservatively chose to estimate the counts in SNR~1885
only from the southern $2 \times 2$ block of binned pixels.

In this $2 \times 2$ block of binned pixels at SNR~1885's position,
the southern 2 of the 4 pixels were contaminated by cosmic rays in each of the
first and third exposures, but the second exposure was clean of cosmic rays.
Thus there are 8 independent measurements of counts in the interior of SNR~1885:
from 2 pixels in each of the first and third exposures,
and from 4 pixels in the second exposure.
The average and dispersion of the 8 measurements was
$9.9 \pm 1.2 \, {\rm DN}$ per binned pixel per exposure,
which at $7.12 \, \cts \, \pix^{-1}$ is
equivalent to $70.4 \pm 8.5 \, {\cts}$ per binned pixel per exposure.
The dispersion of $8.5 \, {\cts}$ is consistent with the
dispersion $\sqrt{70.4} = 8.4$ expected from counting noise.
The uncertainty in the mean counts is $1/\sqrt{8}$ times the dispersion,
that is, $8.5 \, \cts / \sqrt{8} = 3.0 \, \cts$.
Multiplying by 3 to scale to the coadded exposure time
gives a mean count of $211 \pm 9 \, \cts$ per binned pixel
over the combined $8100 \, \s$ exposure.
Subtracting the zero-level of $192 \pm 7 \, \cts$
yields a mean count of $19 \pm 11 \, \cts$ per binned pixel
in the interior of SNR~1885.

The ratio of the net counts $19 \pm 11 \, \cts$ inside SNR~1885
to $79 \pm 29 \, \cts$ in adjacent regions
yields the observed fractional depth of absorption,
$0.24 \pm 0.17$.

\section{Analysis}

The depth of absorption of SNR~1885 observed with the F255W filter,
$0.24 \pm 0.17$,
may be compared to the ratio expected on the basis of a model fit to
the 3240--$4780 \, {\rm \AA}$ absorption line FOS spectrum
reported in \markcite{FGMH}Paper~1.

The model spectrum includes absorption from all non-negligible resonance lines
of neutral and singly-ionized species of
C, O, Mg, Al, Si, S, Ar, Ca,
and iron-group elements with ionic charges from 22 to 30,
namely Ti, V, Cr, Mn, Fe, Co, Ni, Cu, and Zn.
Relative masses of these elements were
set equal to those in the normal SN~Ia model DD21c of
H\"{o}flich et al.\ \markcite{HWT98}(1998).

The adopted ejecta density profile
is the best fit to the \ion{Ca}{1} and \ion{Ca}{2}
absorption line profiles in the FOS spectrum reported in \markcite{FGMH}Paper~1.
The best-fit density profile $n(v)$ is bell-shaped,
a quartic function of free-expansion velocity $v$
up to a maximum velocity $v_{\max} = 13\,100 \, \kms$:
\be
  n(v) \propto \left[ 1 - ( v/v_{\max} )^2 \right]^2
  \ , \quad ( v < v_{\max} )
  \ .
\ee
The available data offer no evidence that the ejecta are compositionally
stratified:
the \ion{Ca}{2} and \ion{Fe}{2} absorption images of SNR~1885
are consistent with being the same size,
and the spectral absorption line profiles of
\ion{Ca}{1}, \ion{Ca}{2}, and \ion{Fe}{1}
are similarly consistent with being the same.
We therefore assume that the compositional structure is fully mixed,
so that all elements follow the same density profile.

\subsection{Photoionization}

As described in \markcite{FGMH}Paper~1,
the freely expanding ejecta in SNR~1885 appear to be in the process of
becoming optically thin to photoionizing radiation,
and are currently undergoing a period of photoionization
by ambient UV starlight out of neutral into the singly-ionized state,
as originally argued by Hamilton \& Fesen \markcite{HF91}(1991).
Recombination is negligible,
with recombination times exceeding a hundred times the age of the remnant.

Calculations of the photoionization of Ca and Fe, the two elements 
observed in the G400H FOS spectrum, were reported in \markcite{FGMH}Paper~1.
Here we complete the account
by reporting photoionization calculations for all elements of interest.

The expected ratios of singly-ionized to neutral species of various elements
depend on how fast they are photoionized out of the assumed initially
neutral state by ambient UV starlight.
Photoionization timescales can be estimated fairly reliably,
at least in the limit where the supernova ejecta are treated as optically thin,
since the spectrum of photoionizing starlight
in the bulge of M31 is observed directly with
{\it IUE\/} (\cite{Burstein88}) and {\it HUT\/} (\cite{FD93}).
Table~\ref{iontable}
lists optically thin photoionization times of all elements considered here,
computed as detailed in \markcite{FGMH}Paper~1.
Photoionization cross-sections were taken from
Verner et al.\ \markcite{Verner96}(1996).
The uncertainty in the photoionization timescales quoted in Table~\ref{iontable}
includes only that arising from uncertainty in the position of SNR~1885
relative to the central nucleus of M31 along the line of sight,
as estimated in \markcite{FGMH}Paper~1,
not uncertainty from the reddening correction or from photoionization
cross-sections.

The photoionization timescales given in Table~\ref{iontable}
are for optically thin ejecta, whereas in fact
the ejecta are expected to be optically thick in broad bands of the ultraviolet,
thanks to resonance line absorption by neutrals and singly-ionized species.
Moreover, the expanding ejecta would have been more optically thick in the past.
An accurate evaluation of the expected ionization structure of SNR~1885
would involve a self-consistent time-dependent computation
of photoionization and radiative transfer in the freely expanding ejecta,
such as was done for SNR~1006 by Hamilton \& Fesen \markcite{HF88}(1988).
However, the present data are too limited,
and the choice of underlying supernova model too uncertain,
to warrant such a computation.

Instead, we estimate the ionization state of different elements
from the optically thin photoionization timescales in Table~\ref{iontable},
together with the observational datum from the FOS spectrum
that calcium is mostly singly ionized, with
$M_\CaII/M_\CaI = 16^{+42}_{-5}$.
According to Table~\ref{iontable},
the optically thin photoionization timescale of \ion{Ca}{1} is
$t_\CaI = 8^{+10}_{-1} \, \yr$.
To reach the observed ionization state of calcium
requires an effective time $t$ given by
$\exp(-t/t_\CaI) = M_\CaI/(M_\CaI+M_\CaII)$,
implying $t = 23^{+30}_{-6} \, \yr$.
In other words,
it is as if calcium has been ionizing not for the full $\sim 110$ year
(at the time of the 1995 FOS observation) age of the remnant, but rather
only for $\sim 20$--$50$ years, because the remnant is only now becoming
optically thin to photoionizing radiation.

If this effective time $t$ is used instead of the age of the remnant,
then the predicted ratio of singly-ionized to neutral species of element Z is
\be
\label{ionfr}
  M_\ZII/M_\ZI = \left( 1 + M_\CaII/M_\CaI \right)^{t_\Ca/t_\Z} - 1
  \ ,
\ee
values of which are given in the third column of Table~\ref{iontable}.
The quoted error on the ratio depends only on the uncertainty in the
observed $M_\CaII/M_\CaI$ ratio,
since uncertainty in the photoionization timescales cancels
in the ratio $t_\Ca/t_\Z$,
to the extent that uncertainties in the relative photoionization cross-sections
are neglected, as here.

\specuvfig

\subsection{Model spectrum}

The upper panel of Figure~\ref{specuv}
shows a model absorption line spectrum
based on the fit to the 3240--$4780 \, {\rm \AA}$
spectrum observed with the G400H grating on the FOS (\markcite{FGMH}Paper~1).
Since singly-ionized to neutral ratios in the best-fit model are skewed
to the low end of the allowed range $M_\CaII/M_\CaI = 16^{+42}_{-5}$
(from which other ionization fractions follow in accordance with
eq.\ [\ref{ionfr}]),
we choose to show not the `best-fit' model,
with $M_\CaII/M_\CaI = 16$,
but rather a `typical' model, with $M_\CaII/M_\CaI = 25$,
the geometric mean of the allowed range $M_\CaII/M_\CaI = 11$--58.
This spectrum is similar
(differing in the adopted ionization fractions)
to the model spectrum shown in Figure~4 of
\markcite{FGMH}Paper~1,
plotted there over the extended range 900--$5000 \, {\rm \AA}$.
The range 1900--$3300 \, {\rm \AA}$
covered in Figure~\ref{specuv} here
includes resonances lines from
\ion{Mg}{1}, \ion{Mg}{2},
\ion{Al}{1},
\ion{Si}{1},
\ion{V}{1}, \ion{V}{2},
\ion{Cr}{2},
\ion{Ti}{2},
\ion{Mn}{1}, \ion{Mn}{2},
\ion{Fe}{1}, \ion{Fe}{2},
\ion{Ni}{1},
and \ion{Zn}{2},
although the contributions from
\ion{Al}{1}, \ion{Si}{1}, and \ion{Ti}{2}
are negligible.

Starlight to the foreground of SNR~1885 is not absorbed.
The fraction of foreground starlight was measured in \markcite{FGMH}Paper~1
from the depth and shape of the \ion{Ca}{2} H \& K lines to be
$0.21^{+0.06}_{-0.12}$.
The model shown in Figure~\ref{specuv} uses the best-fit
foreground starlight fraction of $0.21$.

While the ionization fractions in the model spectrum are fixed
by the observed ionization state of Ca,
and the relative masses of elements are fixed by
H\"{o}flich et al.'s \markcite{HWT98}(1998) theoretical SN~Ia model DD21c,
the overall depth of absorption is scaled so that the depth of \ion{Fe}{1}
absorption is as observed in the FOS spectrum.
The corresponding mass of neutral iron is $M_\FeI = 0.013 \, \Msun$.
At the level of ionization adopted in Figure~\ref{specuv},
the ionization state of iron is $M_\FeII/M_\FeI = 14$,
for a total Fe mass of $M_\FeI + M_\FeII = 0.20 \, \Msun$.

The lower panel of Figure~\ref{specuv}
shows the expected spectrum from the bulge of M31
as measured by {\it IUE\/} (\cite{Burstein88}),
normalized to the surface brightness at the position of SNR~1885,
and multiplied by the throughput of the F255W filter.
The lower panel of Figure~\ref{specuv}
also shows the expected filtered and absorbed spectrum
at the position of SNR~1885, which is the filtered bulge spectrum
multiplied by the absorption curve in the top panel.
The ratio of the area under the SNR~1885 spectrum
to the area under the bulge spectrum
is the predicted fractional depth of absorption,
the expected ratio of counts inside to outside SNR~1885 in the F255W image.

The ratio of counts inside to outside SNR~1885
in the model shown in Figure~\ref{specuv} is $0.33$,
consistent with the observed ratio of $0.24 \pm 0.17$.
Increasing the level of ionization deepens the absorption slightly:
for ionization states varying over $M_\CaII/M_\CaI = 16^{+42}_{-5}$,
the fractional depth of absorption varies over $0.35^{-0.05}_{+0.02}$,
again consistent with the observed ratio of $0.24 \pm 0.17$.

The model level of absorption can be deepened, but only slightly,
from $0.33$ to $0.30$, by reducing the foreground starlight fraction
below the best-fit value of $0.21$,
and at the same time reducing the \ion{Ca}{2} mass
in order to maintain consistency with the depth of the \ion{Ca}{2} H \& K line
profiles observed in the FOS spectrum.

While \ion{Fe}{2} is the main contributor to the expected absorption
in the F255W filter,
\ion{Fe}{1} also makes a significant contribution,
and \ion{Mg}{1}, \ion{Mg}{2} and \ion{Mn}{1}
produce appreciable absorption along the red side of the filter.
Because the \ion{Fe}{2} absorption is heavily saturated,
increasing the amount of \ion{Fe}{2} changes little the
fractional depth of absorption predicted by the model.
The contribution of \ion{Fe}{1} cannot be changed,
since it is tied to the level of absorption observed with the FOS.
Changing the abundance of Mg or Mn on the other hand does have some effect.
For example,
increasing the abundance of Mg by a factor of 5, which is plausible,
deepens the fractional depth of absorption to $0.29$;
increasing Mg to the point where the Mg lines are fully saturated
deepens the fractional depth to $0.26$.

We therefore conclude that the model predicts a fractional depth
of absorption in the rather narrow range $0.33 \pm 0.04$.
While the observed level of UV absorption, $0.24 \pm 0.17$,
is consistent with expectation, it does not constrain strongly the amount
of \ion{Fe}{2} in the ejecta of SN~1885.

\section{Summary}

\ion{Fe}{2} imaging of the remnant of SN~1885 using the F255W filter
on the WFPC2 reveals a dark spot of absorption,
with position and diameter in accord with those measured from
the higher quality WFPC2 \ion{Ca}{2} absorption image
from \markcite{FGMH}Paper~1.

The measured ratio of flux inside to outside SNR~1885 in the \ion{Fe}{2} image
is $0.24 \pm 0.17$,
in good agreement with the ratio $0.33 \pm 0.04$
expected on the basis of a model fit to the near-UV FOS spectrum
reported in \markcite{FGMH}Paper~1.

The observed depth of the \ion{Fe}{2} absorption in SNR~1885 is
consistent with \ion{Fe}{2}
being fully saturated, so that the present data constrain the mass
of iron in the supernova ejecta only weakly.
In particular, the iron mass is consistent with the range
$M_\Fe = 0.1$--$1.0 \, \Msun$ inferred in \markcite{FGMH}Paper~1.
Figure~\ref{specuv} of the present paper illustrates a model UV spectrum
with $M_\Fe = 0.20 \, \Msun$.
The Figure indicates that, besides iron,
ion species \ion{Mg}{1}, \ion{Mg}{2}, and \ion{Mn}{1}
probably make some contribution to the absorption in the F255W image.

Finally, the observed depth of \ion{Fe}{2} absorption
is consistent with the theoretical expectation that the remnant of SN~1885
should have a rich UV spectrum of broad absorption lines.
Unfortunately, the faintness of the bulge of M31 in the UV means that the
signal-to-noise ratio currently attainable with STIS on {\it HST\/} is marginal.
If a UV spectrum with adequate S/N ratio could be obtained,
then it should be possible to constrain the mass and velocity distribution of
Mg,
Si,
Ca,
V,
Cr,
Mn,
Fe,
Co,
Ni,
Cu,
and Zn
in the ejecta of SN~1885.
From such observations it would be possible to learn a great deal 
not only about SN~1885 itself, but also about the increasingly 
important class of subluminous SN~Ia in general
(\cite{Fil97}).

\acknowledgments

We thank
R. McCray for helpful conversations,
and K. McLin for help with data reduction.
RAF is grateful for support from a JILA Visiting Fellowship.
Support for this work was provided by NASA through grant number
GO-6434 from the Space Telescope Science Institute,
which is operated by AURA, Inc., under NASA contract NAS 5-26555.

\iontable


%
%
%
%


\begin{thebibliography}{}

\bibitem[Bertola et al.\ 1995]{Bertola95}
	Bertola, F., Bressan, A., Burstein, D., Buson, L. M., Chiosi, C.,
	\& di Serego Alighieri, S. 1995, \apj, 438, 680
\bibitem[Brown et al.\ 1998]{Brown98}
	Brown, T. M., Ferguson, H. C., Stanford, S. A., \& Deharveng, J. M.
	1998, \apj, 504, 113
\bibitem[Burstein et al.\ 1988]{Burstein88}
	Burstein, D., Bertola, F., Buson, L. M., Faber, S. M., \& Lauer, T. R.
	1988, \apj, 328, 440
\bibitem[de Vaucouleurs \& Corwin 1985]{DVC85}
	de Vaucouleurs, G., \& Corwin, H.  G. Jr. 1985, \apj, 295, 287 (dVC)
\bibitem[Ferguson \& Davidsen 1993]{FD93}
	Ferguson, H. C., \& Davidsen, A. F. 1993, \apj, 408, 92
\bibitem[Fesen, Hamilton, \& Saken 1989]{FHS}
	Fesen, R. A., Hamilton, A. J.  S., \& Saken, J. M. 1989,
	\apj, 341, L55
\bibitem[Fesen et al.\ 1999]{FGMH}
	Fesen, R. A., Gerardy, C. L., McLin, K. M., \& Hamilton, A. J. S.
	1999, \apj, 514, 195 (astro-ph/9810002) (Paper~1)
\bibitem[Filippenko et al.\ 1992]{Fil92}
	Filippenko, A. V., et al.\ 1992, \aj, 104, 1543
\bibitem[Filippenko 1997]{Fil97}
	Filippenko, A. V. 1997, \araa, 35, 309
\bibitem[Hamilton \& Fesen 1988]{HF88}
	Hamilton, A. J. S., \& Fesen, R. A.  1988, \apj, 327, 178
\bibitem[Hamilton \& Fesen 1991]{HF91}
	Hamilton, A. J. S., \& Fesen, R. A.  1991, in Supernovae,
	10th Santa Cruz Summer Workshop in Astronomy and Astrophysics,
	ed.\ S. E. Woosley (Berlin: Springer-Verlag), 656
\bibitem[H\"{o}flich \& Khokhlov 1996]{HK96}
	H\"{o}flich, P., \& Khokhlov, A. 1996, \apj, 457, 500
\bibitem[H\"{o}flich, Wheeler \& Thielemann 1998]{HWT98}
	H\"{o}flich, P., Wheeler, J. C., \& Thielemann, F.-K. 1998,
	\apj, 495, 617
\bibitem[Lauer et al.\ 1998]{Lauer98}
	Lauer, T. R., Faber, S. M., Ajhar, E. A., Grillmair, C. J.,
	\& Scowen, P. A. 1998, \aj, 116, 2263
\bibitem[Nomoto, Thielemann, \& Yokoi 1984]{NTY84}
	Nomoto, K., Thielemann, F.-K., \& Yokoi, K. 1984, \apj, 286, 644
\bibitem[Nomoto et al.\ 1997]{N97}
	Nomoto, K., Iwamoto, K., Nakasato, N., Thielemann, F.-K.,
	Brachwitz, F., Tsujimoto, T., Kubo, Y., \& Kishimoto, N. 1997,
	Nuclear Physics A, 621, 467c
\bibitem[O'Connell 1999]{OConnell99}
	O'Connell, R. W. 1999, \araa, to appear
	(astro-ph/9906068)
\bibitem[Tammann \& Reindl 1999]{Tammann99}
	Tammann, G. A., \& Reindl B. 1999,
	Proc.\ Supernova Workshop, Assergi, Sep~29--Oct~2 1998
	(astro-ph/9903220)
\bibitem[van den Bergh 1994]{vdb94}
	van den Bergh, S. 1994, \apj, 424, 345
\bibitem[Verner et al.\ 1996]{Verner96}
	Verner, D. A., Ferland, G. J., Korista, K. T., \& Yakovlev, D. G. 1996,
	\apj, 465, 487
\bibitem[Woosley 1997]{W97}
	Woosley, S. E. 1997, \apj, 476, 801
\bibitem[Woosley \& Weaver 1994]{WW94}
	Woosley, S. E., \& Weaver, T. A. 1994, \apj, 423, 371
\bibitem[Zwicky 1958]{Zwicky}
	Zwicky, F. 1958, Handbuch der Physik, 51, 766 (Berlin: Springer-Verlag)
\end{thebibliography}
\end{document}